\documentclass[amsmath,amssymb,superscriptaddress,aps,twocolumn]{revtex4}

\usepackage{hyperref}
\usepackage{graphicx}
\usepackage{floatrow}

\hypersetup{
	final=true,
	colorlinks=true,
	linkcolor=blue,
	citecolor=blue,
	urlcolor=blue}

\newcommand{\sn}[2]{\ensuremath{{#1}{\times}10^{#2}}} 
\newcommand{\TiSe}{1\textit{T}-TiSe$_{2}$}
\newcommand{\VSe}{1\textit{T}-VSe$_{2}$}
\newcommand{\degsC}{$^{\circ}$C}
\newcommand{\Tcdw}{$T_{\mathrm{CDW}}$}

\newcommand{\Tgrowth}{$T_{\mathrm{g}}$}
\newcommand{\Treact}{$T_{\mathrm{r}}$}
\newcommand{\SeTwo}{Se$^{2-}$}
\newcommand{\SeZero}{Se$^{0}$}
\newcommand{\SeMinus}{Se$^{(-)}$}
\newcommand{\dRdT}{d$R$/d$T$}
\newcommand{\hv}{$h\nu$}
\newcommand{\overbar}[1]{\mkern 1.5mu\overline{\mkern-1.5mu#1\mkern-1.5mu}\mkern 1.5mu}

\begin{document}



\title{Correlation between crystal purity and the charge density wave in 1\textit{T}-VSe$_{2}$}

\date{\today}


\author{C.~J.~Sayers}
\email[Corresponding author: ]{c.j.sayers@bath.ac.uk}
\affiliation{Centre for Nanoscience and Nanotechnology, Department of Physics, University of Bath, Bath, BA2 7AY, UK}

\author{L.~S.~Farrar}
\affiliation{Centre for Nanoscience and Nanotechnology, Department of Physics, University of Bath, Bath, BA2 7AY, UK}

\author{S.~J.~Bending}
\affiliation{Centre for Nanoscience and Nanotechnology, Department of Physics, University of Bath, Bath, BA2 7AY, UK}

\author{M.~Cattelan}
\affiliation{School of Chemistry, University of Bristol, Cantocks Close, Bristol BS8 1TS, UK}

\author{A.~J.~H.~Jones}
\affiliation{HH Wills Physics Laboratory, University of Bristol, Tyndall Avenue, Bristol BS8 1TL, UK}

\author{N.~A.~Fox}
\affiliation{School of Chemistry, University of Bristol, Cantocks Close, Bristol BS8 1TS, UK}
\affiliation{HH Wills Physics Laboratory, University of Bristol, Tyndall Avenue, Bristol BS8 1TL, UK}

\author{G.~Kociok-K{\"o}hn}
\affiliation{Material and Chemical Characterisation Facility (MC$^{2}$), University of Bath, Claverton Down, Bath BA2 7AY, UK}

\author{K.~Koshmak}
\affiliation{IOM-CNR Institute, Area Science Park, SS 14 Km, 163.5, Basovizza, 34149 Trieste, Italy}

\author{J.~Laverock}
\affiliation{HH Wills Physics Laboratory, University of Bristol, Tyndall Avenue, Bristol BS8 1TL, UK}

\author{L.~Pasquali}
\affiliation{IOM-CNR Institute, Area Science Park, SS 14 Km, 163.5, Basovizza, 34149 Trieste, Italy}
\affiliation{Dipartimento di Ingegneria ``Enzo Ferrari", Universit{\`a} di Modena e Reggio Emilia, via P. Vivarelli 10, 41125 Modena, Italy}
\affiliation{Department of Physics, University of Johannesburg, P.O. Box 524, Auckland Park 2006, South Africa}

\author{E.~Da Como}
\affiliation{Centre for Nanoscience and Nanotechnology, Department of Physics, University of Bath, Bath, BA2 7AY, UK}

\begin{abstract}
	We examine the charge density wave (CDW) properties of 1\textit{T}-VSe$_{2}$ crystals grown by chemical vapour transport (CVT) under varying conditions. Specifically, we find that by lowering the growth temperature (\textit{T}$_{\mathrm{g}}$~$<$ 630$^{\circ}$C), there is a significant increase in both the CDW transition temperature and the residual resistance ratio (RRR) obtained from electrical transport measurements. Using x-ray photoelectron spectroscopy (XPS), we correlate the observed CDW properties with stoichiometry and the nature of defects. In addition, we have optimized a method to grow ultra-high purity 1\textit{T}-VSe$_{2}$ crystals with a CDW transition temperature, $T_{\mathrm{CDW}}$ = (112.7 $\pm$ 0.8) K and maximum residual resistance ratio (RRR) $\approx$ 49, which is the highest reported thus far. This work highlights the sensitivity of the CDW in 1\textit{T}-VSe$_{2}$ to defects and overall stoichiometry, and the importance of controlling the crystal growth conditions of strongly-correlated transition metal dichalcogenides.
	
\end{abstract}


\maketitle


\section{Introduction}\label{Sec:Introduction}

\begin{figure}[ht!]
	\begin{center}
		\includegraphics[width=1.0\linewidth,clip]{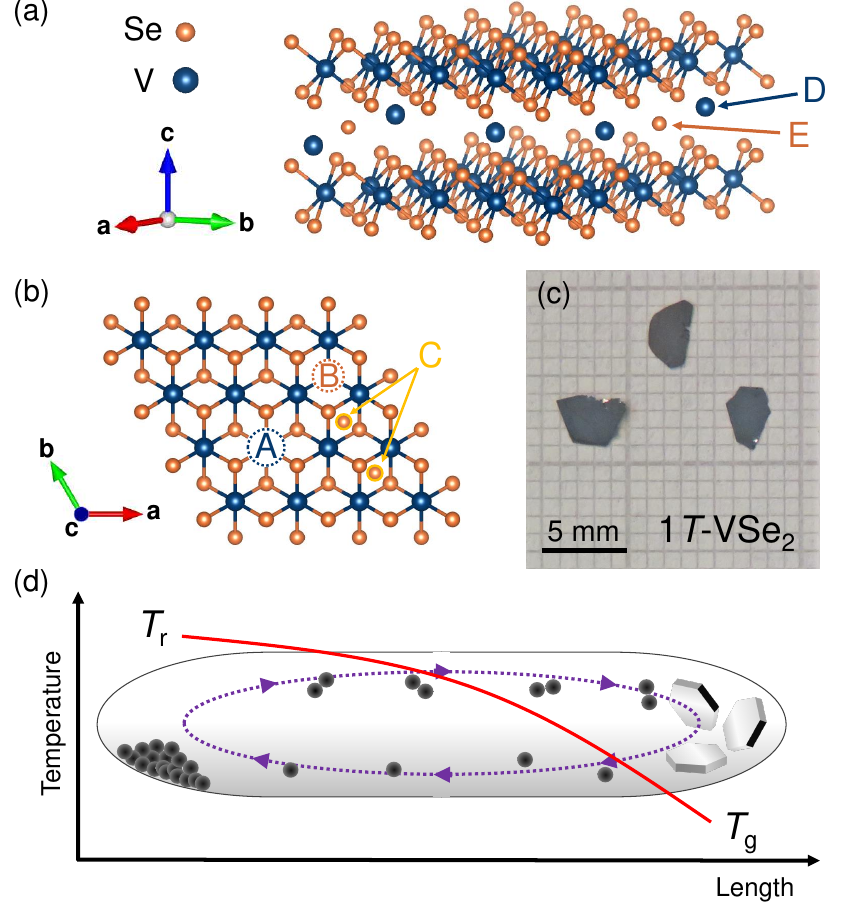}
	\end{center}
	\vspace*{-0.2in}
	\caption{\label{fig:Growth}~Growth of \VSe~crystals.~Panels (a) and (b) show the crystal structure of \VSe~viewed along the layer plane and $c$-axis respectively. Common defect types found in TMDs are labelled. In \VSe, these correspond to vanadium and selenium vacancies (A and B), selenium aggregates/clusters (C), and interstitials in the van der Waals (vdW) gap between the layer planes (D and E). (c) Image of typical crystals produced by CVT (d) Illustration of the chemical vapour transport (CVT) process. Iodine vapour carries material from the reaction zone at \Treact~to the growth zone at \Tgrowth~where crystals form. The solid red line is a sketch of a typical temperature profile.}
\end{figure}

The metallic transition metal dichalcogenides (TMDs) are well-known to exhibit interesting strongly correlated behaviour such as charge density waves (CDWs) \cite{Wilson1975,Rossnagel2011} and superconductivity. In \VSe, an incommensurate CDW develops at \Tcdw~= 110 K with (4\textit{a} x 4\textit{a} x 3.18\textit{c}) periodic lattice distortion, followed by a further change in the distorted $c$-axis to 3.25\textit{c} below 85 K \cite{Tsutsumi1982}. Signatures of these phase transitions have been noted in x-ray \cite{Tsutsumi1982} and electron diffraction \cite{Eaglesham1986}, magnetic and transport studies \cite{Thompson1979}. The CDW is likely driven by Fermi surface nesting involving states along the flat portions of the electron pocket centred around the $\overbar{M}$-point at the edge of the Brillouin zone \cite{Terashima2003}, which possibly has a three-dimensional character \cite{Strocov2012}. The common polytype of VSe$_{2}$ has the trigonal (1\textit{T}) unit cell and belongs to the P$\overbar{3}$m1 spacegroup \cite{Eaglesham1986}. Fig.~\ref{fig:Growth}(a) and (b) shows the layered structure consisting of Se-V-Se planes in the $a$-$b$ direction seperated by a van der Waals (vdW) gap, and octahedral metal coordination when viewed along the $c$-axis.

In addition to the ongoing efforts to understand CDW transitions in bulk TMDs, the recent surge in research of layered 2D materials has reignited interest in this area. So far, it has been shown that \VSe~has a large range of tunability of its CDW properties with dimensionality, as the transition temperature and magnitude of the order parameter are strongly influenced by the sample thickness on a nanometre scale \cite{Pasztor2017,Yang2014}. Control of the distorted lattice periodicity is also possible by strain engineering, and an unconventional (4\textit{a} x $\sqrt{3}$\textit{a}) CDW has been observed by STM \cite{Zhang2017}. At the monolayer limit, there have been reports of possible ferromagnetism \cite{Ma2012,Bonilla2018,Liu2018}, strongly enhanced CDW order evidenced by a fully-gapped Fermi surface \cite{Feng2018}, and an increase of the transition temperature up to \Tcdw~= 220 K  \cite{Chen2018}.

With increasing interest in studying monolayer or few-layer charge density wave TMDs, the bottom-up approach of growth by molecular beam epitaxy (MBE) \cite{Zhang2017,Feng2018} or chemical vapour deposition (CVD) \cite{Fu2016,Wang2017} techniques has become common. Once optimized, these methods are excellent at producing thin films. However, samples produced this way often suffer from a high number of crystal defects. In particular, the presence of grain boundaries due to coalescing nucleation sites is a common and well-documented problem in 2D materials grown by CVD, such as graphene \cite{Yu2011}. Also, there is often an unavoidable interaction with the growth substrate such as doping \cite{Peng2015} or strain \cite{Zhang2017} which can alter the sample properties. Instead, a top-down approach by exfoliation of high-quality crystals as the starting material is still considered the ideal way to obtain pristine monolayers and to provide the basis for construction of van der Waals heterostructures by dry transfer \cite{Castellanos-Gomez2014}.

At present, many researchers either grow their own crystals or obtain them commercially. In both cases, the specific growth conditions are often not reported. However, notably among the TMDs, \VSe~can grow far from ideal stoichiometry in varying conditions \cite{Hayashi1978}. Previous investigations on \VSe~have suggested that the CDW transition temperature is maximum for samples prepared using the lowest growth temperatures (\Tgrowth~$<$ 600\degsC) where the Se:V ratio is expected to approach 2:1 \cite{Levy1979,DiSalvo1981}. Deviations from ideal stoichiometry due to increasing density of defects can occur at higher growth temperatures, but their nature in this compound remains unknown. Types of defects that are typically found in TMD crystals are labelled in Fig.~\ref{fig:Growth}(a) and (b) including vacancies and interstitials \cite{Hildebrand2014}. The influence of growth temperature on crystal purity and its relation to CDW properties has been reported in the related compound 1\textit{T}-TiSe$_{2}$. Increasing Ti interstitials in the vdW gap between the layers results in electron doping and reduces \Tcdw~by upsetting the delicate balance of electrons to holes which has consequences within the excitonic insulator scenario \cite{DiSalvo1976}. In \VSe, where Fermi surface nesting is suggested to play a role, such a doping effect could have a significant impact on the CDW by altering the topology of the Fermi surface due to the change in chemical potential. Thus far, a detailed study of the growth conditions of \VSe~in relation to its CDW properties and the role of defects is lacking. 

Here, we show that changes in the growth temperature over the range (550 - 700)\degsC~have a significant impact on the physical purity, structure, and intrinsic stoichiometry which subsequently influences the CDW behaviour. Electronic transport measurements show that the CDW transition temperature increases markedly for crystals grown with \Tgrowth~$<$ 630\degsC. Using x-ray spectroscopy, we are able to identify defects in samples grown at higher temperatures, and correlate this with the observed electronic properties. In addition, we have optimized the conditions for growing ultra-pure \VSe~at \Tgrowth~= 550\degsC~with a transition temperature of \Tcdw~= (112.7 $\pm$ 0.8) K and a maximum residual resistance ratio (RRR) $\approx$ 49 which, to the best of our knowledge, surpasses the highest value reported previously of $\sim$28 \cite{Toriumi1981,Yadav2010}.

\section{Experimental methods}

\subsection{Crystal growth}\label{Sec:methods, crystal growth}

\VSe~crystals were grown using a chemical vapour transport (CVT) method \cite{Hedayat2019}, whereby a constant temperature gradient drives the crystallisation of the elements from a vapour at high temperature as illustrated in Fig.~\ref{fig:Growth}(d). High purity vanadium (99.9\%) and selenium (99.99\%) powders were sealed inside an evacuated quartz ampoule, together with anhydrous iodine (99.998\%) which acts as the transport agent. Under inert atmosphere (N$_{2}$ glovebox), stoichiometric amounts of the elements (2:1 molar ratio) were loaded plus a slight excess of selenium (2.4 - 3.4 mg~cm$^{-3}$), based on the ampoule volume. The amount of iodine (2.10 - 2.25 mg~cm$^{-3}$) was chosen to facilitate growth within a reasonable timeframe \cite{Rimmington1974}, whilst also minimizing the potential for iodine to be introduced into the crystals which is known to suppress the CDW transition in other TMDs \cite{DiSalvo1976}. The ampoule was evacuated using a pump capable of achieving base pressure (\sn{7}{-6}) mbar and then sealed with a flame. A two-zone tube furnace was used with a temperature gradient between hot (reaction temperature, \Treact) and cold (growth temperature, \Tgrowth) zones of $\Delta T$ = (55 - 60)\degsC~which was confirmed with a thermocouple probe. The temperature stability was better than $\pm$ 1\degsC. Growth proceeded for up to 21 days, resulting in many crystals of (5 x 5 x 0.1 mm) forming at the cold end of the ampoule. The crystals were typically thin platelets with hexagonal edges and a highly reflective silver appearance as shown in Fig.~\ref{fig:Growth}(c).

With these general conditions, this procedure was carried out several times in order to produce separate sample batches with different growth temperatures, \Tgrowth~in the range (550 - 700)\degsC. 

\subsection{Electronic transport}\label{Sec:methods, Electronic transport}

We measured the resistance as a function of temperature, R(\textit{T}) for different crystal batches. Samples were cut with a razor blade into rectangular shapes (typical lateral size 4 mm x 1 mm) and mounted on electrically insulating substrates using thermal varnish. Contacts were made to the as-grown crystal surfaces using silver paste in a standard 4-point configuration. Measurements were performed in a JANIS 4 K closed cycle cryocooler (4 - 300 K) using a lock-in amplifier (SR830) and a typical excitation current of 1 mA.

Multiple crystals were measured from each batch to investigate any possible variations.

\subsection{X-ray spectroscopy}\label{Sec:methods, X-ray spectroscopy}

We used both x-ray photoelectron spectroscopy (XPS) and x-ray absorption spectroscopy (XAS) with synchrotron radiation to study core levels of selenium (Se), vanadium (V) and iodine (I) in \VSe. Measurements were performed at the BEAR end station (Elettra synchrotron, Trieste) \cite{Nannarone2004,Pasquali2004}. In order to expose a clean surface, the samples were cleaved in a nitrogen glove-bag attached to the fast entry of the preparation chamber. 

We found that after prolonged exposure, we were able to induce radiation damage to the sample surface which resulted in noticeable changes in the spectra. Hence, for the results presented in the main text and the Supplemental Material (Ref.~\cite{Supp}), we minimized both the exposure time and photon flux (e.g. $\sim$ \sn{7}{10} photons s$^{-1}$ at \hv~=150 eV). Identical experimental conditions were maintained for each sample. As a result of these measures, we did not observe any evolution of the spectra during the measurements which could be related to damage and therefore the properties we report are intrinsic to the samples.

\section{Electronic Properties}\label{Sec:Electronic properties}

\begin{figure}[ht!]
	\begin{center}
		\includegraphics[width=1.0\linewidth,clip]{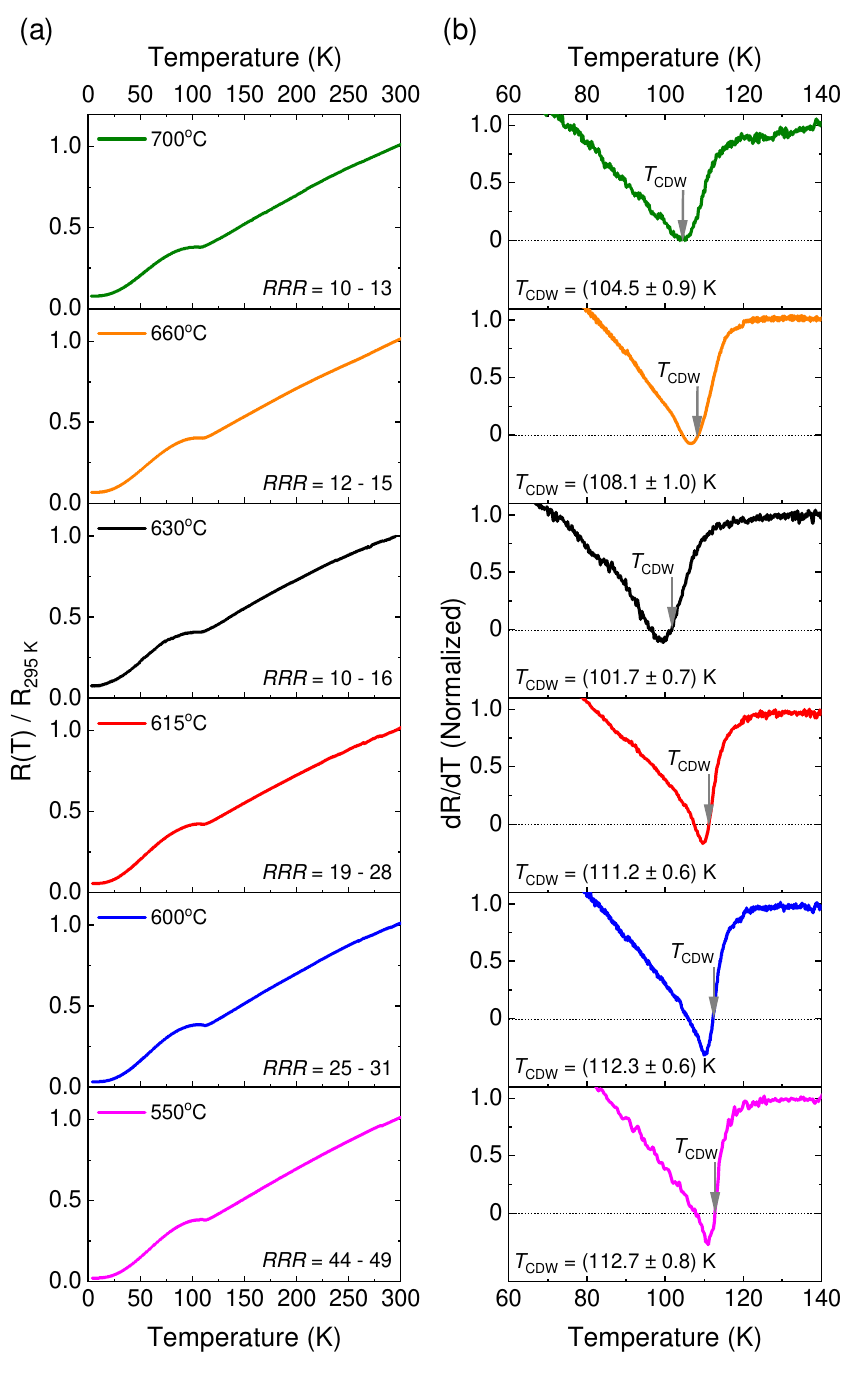}
	\end{center}
	\vspace*{-0.2in}
	\caption{\label{fig:Transport1} Resistance of \VSe~samples produced with different growth temperatures as indicated. (a) Normalized resistance, R/R$_{\mathrm{295 K}}$ for different sample batches. The range of residual resistance ratio (RRR) for each crystal batch is labelled. (b) First derivative of resistance, \dRdT~highlighting the charge density wave transition, \Tcdw~which we define as the point at which \dRdT~initially crosses zero indicated by the arrows.}
\end{figure}

\begin{figure}[h!]
	\begin{center}
		\includegraphics[width=1.0\linewidth,clip]{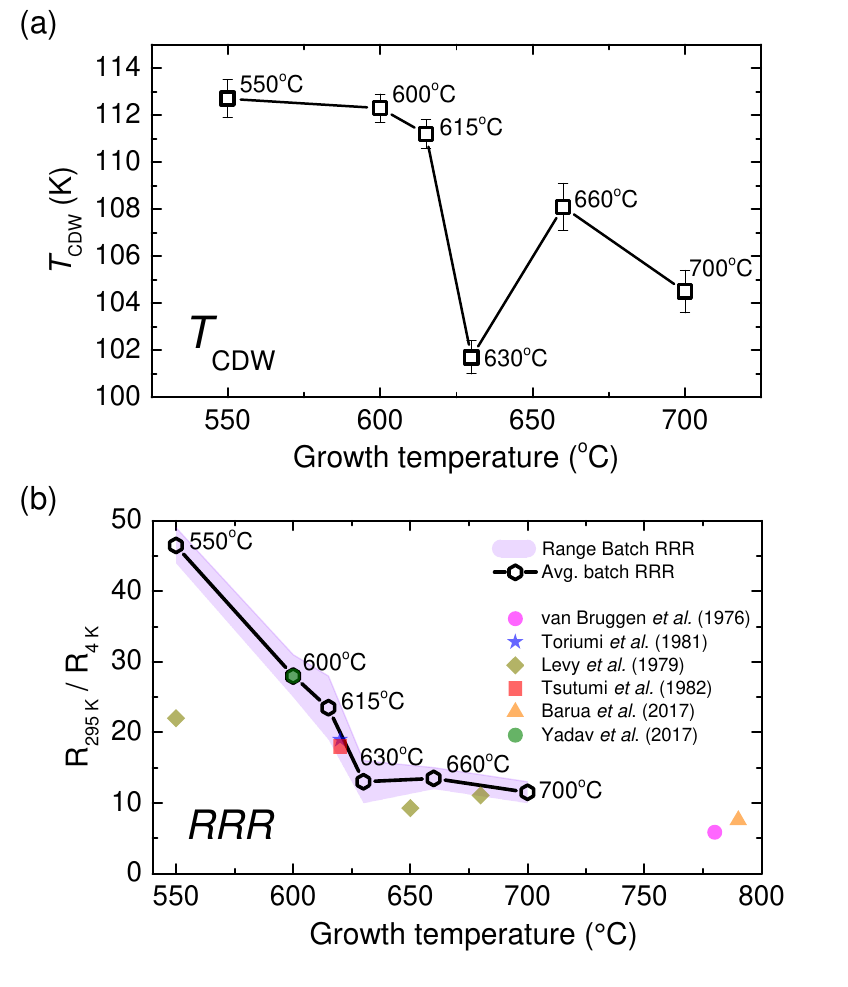}
	\end{center}
	\vspace*{-0.2in}
	\caption{\label{fig:Transport2} Electronic properties of \VSe~samples as a function of crystal growth temperature, \Tgrowth~(a) Charge density wave transition temperature, \Tcdw. (b) Average residual resistance ratio (RRR), R$_{\mathrm{295 K}}$/R$_{\mathrm{4 K}}$. The shaded area corresponds to the RRR range found within each batch of crystals. For comparison, the coloured data points show RRR values taken from the literature as indicated. If the value was not directly quoted in those works, it has been estimated here from the published resistance/resistivity data.}
\end{figure}

To investigate the bulk electronic properties of \VSe, we measured the resistance as a function of temperature, R(\textit{T}) for different crystal batches as shown in  Fig.~\ref{fig:Transport1}(a). The shape of the curve agrees with previous studies of \VSe~and shows an overall metallic behaviour with a slight increase in the resistance related to the CDW transition at \Tcdw~$\approx$ (100 - 110) K \cite{vanBruggen1976,Bayard1976,Thompson1979}. At low temperature, the resistivity approaches a finite value related to the amount of impurity scattering \cite{vanBruggen1976}. We do not see any indication of an upturn in the resistance of our samples at low temperature (4 K) in contrast to a previous study, which attributed this behaviour to a Kondo effect \cite{Barua2017}.

The first derivative, \dRdT~in Fig.~\ref{fig:Transport1}(b) highlights the CDW transition, \Tcdw~which we define as the onset temperature where the resistance first starts to increase, and corresponds to the point at which \dRdT~initially crosses zero [dashed horizontal line in Fig.~\ref{fig:Transport1}(b)]. Details of obtaining \Tcdw~in this way are provided in Ref.~\cite{Supp}, and we emphasize that by using alternative points in \dRdT~to define \Tcdw~does not change the observed trend between samples or the conclusions of this work. Fig.~\ref{fig:Transport2}(a) shows the CDW transition temperatures extracted from \dRdT~for different sample batches as a function of growth temperature. The overall trend shows that \Tcdw~is increased at lower growth temperatures, reaching a maximum of \Tcdw~= (112.7 $\pm$ 0.8) K for the 550\degsC~sample. We note that samples with \Tgrowth~$<$ 630\degsC~show an increase in transition temperature above the typically reported value of \Tcdw~$=$ 110 K. Based on the trend of our samples produced at the lowest growth temperatures, it seems as if \Tcdw~is approaching the inherent maximum for \VSe. It is also apparent in Fig.~\ref{fig:Transport1}(b) that there is an evolution in the width of \dRdT~feature relating to the CDW transition, where there is a broadening at higher growth temperatures. This is seen in both the onset of the transition (gradient of the steep decline near \Tcdw)~and width of the minima in \dRdT. Such a broadening is often linked to disorder, and the behaviour in our samples is similar to that observed in NbSe$_{2}$ using electron irradiation dosing to induce atomic defects without doping the system \cite{Cho2018}. Additional information can be extracted from the magnitude of the \dRdT~minima below zero (dashed horizontal line) which also follows a similar trend, as \dRdT~of the sample with highest \Tcdw~becomes the most negative following the CDW transition. We show in Ref.~\cite{Supp} that this corresponds to the magnitude of the resistance increase at \Tcdw~relative to the normal metallic behaviour. Therefore our results are fully consistent with strengthened CDW order where one would expect a more effective gapping of the Fermi surface, as the proportion of charge carriers removed at \Tcdw~is greater.

The residual resistance ratio (RRR) is a well-known parameter often used to estimate the purity of metals. At sufficiently low temperatures, the scattering of carriers is dominated by impurities or crystal defects, which gives rise to a finite (residual) resistance. The magnitude of this residual resistance is thus linked to the density of impurities and/or defects present in the sample. Using the ratio R$_{\mathrm{295 K}}$/R$_{\mathrm{4 K}}$, we are able to compare several samples as shown in Fig.~\ref{fig:Transport2}(b). Overall, we find that the average RRR within each batch increases for lower growth temperatures, indicating a reduction in impurity/defect scattering. For samples grown at the lowest temperature (550\degsC), we find a maximum RRR of $\sim$ 49 within the batch. Similar to the behaviour of the CDW transition temperature, there is also a significant increase in the average RRR from $\sim$ 12 to $>$ 22 for samples with \Tgrowth~$<$ 630\degsC. This is consistent with the width of the CDW transitions in \dRdT~[Fig.~\ref{fig:Transport1}(b)] for samples \Tgrowth~= (630 - 700)\degsC, confirming a defect-induced broadening. By comparing our results to reports of RRR in the literature on \VSe~over a range of growth temperatures, there is a clear agreement with this trend [Fig.~\ref{fig:Transport2}(b)].

Finally, we discuss the scattering mechanism of charge carriers at low temperature by analyzing the temperature dependence of the resistance. Previously in \VSe, there have been reports of a strong \textit{T}$^{2}$ dependence \cite{Bayard1976,Thompson1979} which could suggest a significant electron-electron scattering. By fitting the low temperature data in the range (5 - 15) K with a power law, R $\propto T^{x}$ we also find a \textit{T}$^{2}$ dependence but only for the sample grown at the highest temperature (700\degsC). By contrast, the sample grown at the lowest temperature (550\degsC) shows a \textit{T}$^{3}$ behaviour, and there appears to be a progressive shift between these two regimes for intermediate temperatures (see Ref.~\cite{Supp}). In other CDW-bearing TMDs, a temperature dependence in the range \textit{T}$^{3}$ - \textit{T}$^{5}$ is found for $T \leq$ 25 K, depending on the compound, and is related to electron-phonon scattering \cite{Naito1982}. The power then depends on the orbital character of bands involved in either an \textit{intraband} or \textit{interband} scattering, the density of states, and strength of the CDW order (proportion of carriers lost from the Fermi surface). For example, a \textit{T}$^{3}$ dependence is found for the ideal resistivity in 2\textit{H}-NbSe$_{2}$ and can be understood by a simple two-band model which describes an electron-phonon mediated \textit{interband} scattering of $s$-like electrons into the $d$-band. Such a two-band analysis has also been used to successfully describe the temperature dependence of the Hall coefficient in 1\textit{T}-VSe$_{2}$ \cite{Toriumi1981} and the ideal resistivity was found to be in the range \textit{T}$^{3}$ - \textit{T}$^{4}$ for the highest purity samples.

We therefore suggest that a similar electron-phonon scattering mechanism is the case for pure \VSe~which is seen here in our sample with largest RRR (550\degsC). Instead, the low temperature resistance deviates rapidly from this behaviour with increasing disorder and approaches a \textit{T}$^{2}$ dependence. Clearly, this behaviour seems to be rather sample dependent and could explain the large variation reported in the literature \cite{Bayard1976,Thompson1979,Toriumi1981,Yadav2010}

\begin{figure*}[ht!]
	\begin{center}
		\includegraphics[width=0.9\linewidth,clip]{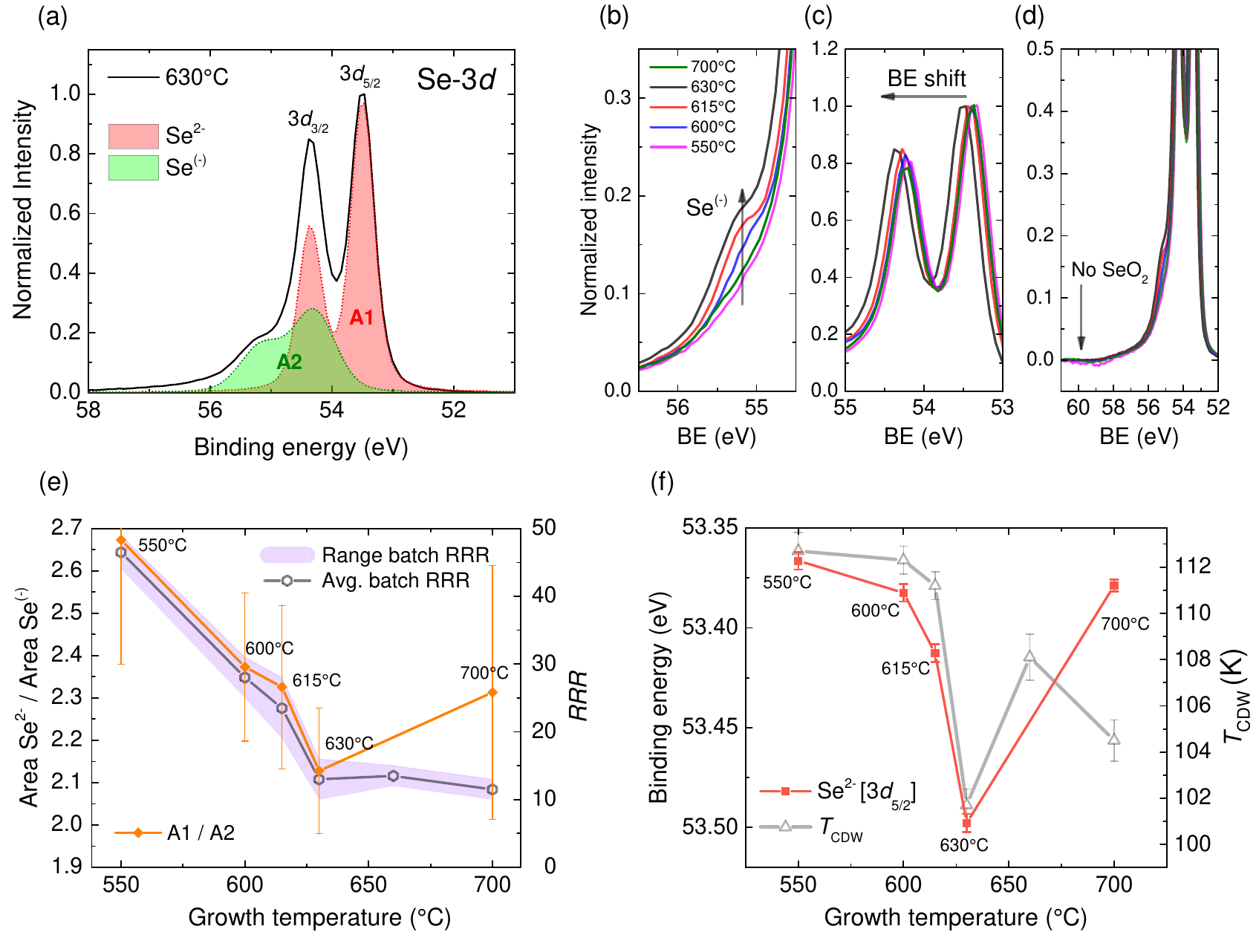}
	\end{center}
	\vspace*{-0.2in}
	\caption{\label{fig:Se-3d} X-ray photoelectron spectroscopy (XPS) of \VSe~samples grown at different temperatures. (a) Se-3\textit{d} spectra (\hv~= 150 eV). The shaded curves are fits to the data for the \Tgrowth~= 630\degsC~sample, showing the contribution from Se$^{2-}$ and Se$^{(-)}$ oxidation states. Panels (b) - (d) show the same data as panel (a) and highlight the feature in Se-3\textit{d} spectra related to the presence of Se$^{(-)}$, the observed shift in binding energy (BE), and the absence of SeO$_{2}$ (e) The area ratio (A1/A2) as a function of growth temperature, \Tgrowth~(left axis) obtained from fitting the Se-3\textit{d} spectra for the Se$^{2-}$ (A1) and Se$^{(-)}$ (A2) oxidation states. A comparison (right axes) is made with the residual resistance ratio (RRR) from electronic transport measurements. (f) Binding energy of the 3\textit{d}$_{5/2}$ peak for Se$^{2-}$ obtained from fitting the data. Error bars in panels (e) and (f) is the standard deviation on the fitting parameters.}
\end{figure*}

In summary, the electronic properties of \VSe~and its CDW behaviour vary quite significantly with crystal growth temperature. The observed trend of \Tcdw~and RRR in Fig.~\ref{fig:Transport2} are similar and therefore, the underlying mechanism influencing both of these properties is likely to be intrinsically linked. However, at low growth temperatures (550 - 600\degsC), it appears that the strength of CDW order (\Tcdw) approaches its maximum, whereas the crystal purity (RRR) continues to increase.

\section{X-ray spectroscopy \& discussion}\label{Sec:XPS}

To obtain information about the chemical composition of our samples, we used XPS to study core levels in \VSe. Herein, we will mainly discuss the XPS results of the Se-3\textit{d} core levels for which we find a significant variation between our samples. Instead, the vanadium core levels show no noticeable variation and iodine was either not present in our samples or its concentration was below the detectable limit. These results are included in the Supplemental Material (Ref.~\cite{Supp}). We did not observe any sign of oxidation in the Se-3\textit{d} spectra which otherwise would be in the form of SeO$_{2}$ present at $\sim$ 59.9 eV \cite{Shenasa1986} (indicated by the arrow in Fig.~\ref{fig:Se-3d}(d)). Additionally, the XAS results near the O K-edge showed a negligible amount of oxygen and all samples had identical spectra (see Ref.~\cite{Supp}). Therefore, we rule out the presence of oxygen or iodine as an explanation for the observed change in CDW properties.

XPS spectra for the Se-3\textit{d} core levels of different samples are shown in Fig.~\ref{fig:Se-3d}(a). The main feature is the doublet (labelled) with Se-3\textit{d}$_{5/2}$ $\approx$ 53.4 eV and spin-orbit splitting $\Delta_{\mathrm{so}} \approx$ 0.86 eV, which corresponds to the \SeTwo~oxidation state of Se bound to V in \VSe~\cite{Liu2018}. An additional feature on the high energy side was also found in the region 55 - 56 eV which is highlighted in Fig.~\ref{fig:Se-3d}(b) and shows a clear variation between samples. We find a good fit to the data using two doublets as illustrated in Fig.~\ref{fig:Se-3d}(a) for the \Tgrowth~= 630\degsC~sample. The binding energy (BE) of the second doublet relating to the additional feature is Se-3\textit{d}$_{5/2}$ $\approx$ 54.2 eV, which we assign to the presence of more positive oxidation states of selenium. Since it appears in our data as a single broad doublet (gaussian width $\sim$2 larger than \SeTwo), it is likely to arise from a combination of oxidation states. We label this feature \SeMinus~since its binding energy is slightly lower (i.e. more negative) than pure selenium, \SeZero~which is expected at $\sim$ 55.5 eV \cite{Shenasa1986}. Such a feature could indicate the presence of partially bound or unbound Se in the form of vanadium vacancies, in-plane Se aggregates/clusters \cite{Liu2018,Chia2016} or Se interstitials \cite{Peng2015} (See Fig.~\ref{fig:Growth}). From the fitting, the contributions of \SeTwo~and \SeMinus~states in each sample are given by the shaded areas, A1 and A2 respectively in Fig.~\ref{fig:Se-3d}(a). The ratio A1/A2 gives an indication of the relative amount of \SeMinus~oxidation states present in our samples, and is shown in Fig.~\ref{fig:Se-3d}(e) as a function of growth temperature. Comparing this ratio with the RRR from electronic measurements [right axis, Fig.~\ref{fig:Se-3d}(e)], there appears to be a correlation with the overall trend which suggests that the presence of these \SeMinus~states contribute to the impurity/defect scattering in electronic transport.

It can also be seen in Fig.~\ref{fig:Se-3d}(c) that there is a shift in binding energy of the Se-3\textit{d} spectra between samples. We find that this shift is significant ($\sim$ 120 meV variation), and there is an overall trend of increasing \SeTwo~binding energy for samples grown at higher temperatures as shown in Fig.~\ref{fig:Se-3d}(f). We rule out any energy shift due to local charge effects from increasing concentration of the more positive \SeMinus~oxidation states discussed previously, since there is negligible broadening of the observed \SeTwo~linewidth between samples. Such a local effect would rather be expected to contribute multiple environments, leading to the inhomogeneous broadening of the observed core level in addition to a binding energy shift. Instead, the rigid shift of the entire Se-3\textit{d} states is better explained by a doping effect in our samples. Specifically, an overall electron doping raises the Fermi level resulting in an apparent shift of core levels to higher binding energies with respect to our XPS energy analyzer. Therefore, we suggest that samples grown at higher temperatures are either V-rich (V$_{1+x}$Se$_{2}$) or Se-deficient (VSe$_{2-x}$). The binding energy shift agrees with the evolution of \Tcdw~[right axis in Fig.~\ref{fig:Se-3d}(f)], and hence we suggest that the suppression of charge density wave order in \VSe~could be related to effective electron doping.

\begin{figure}[ht!]
	\begin{center}
		\includegraphics[width=1.0\linewidth,clip]{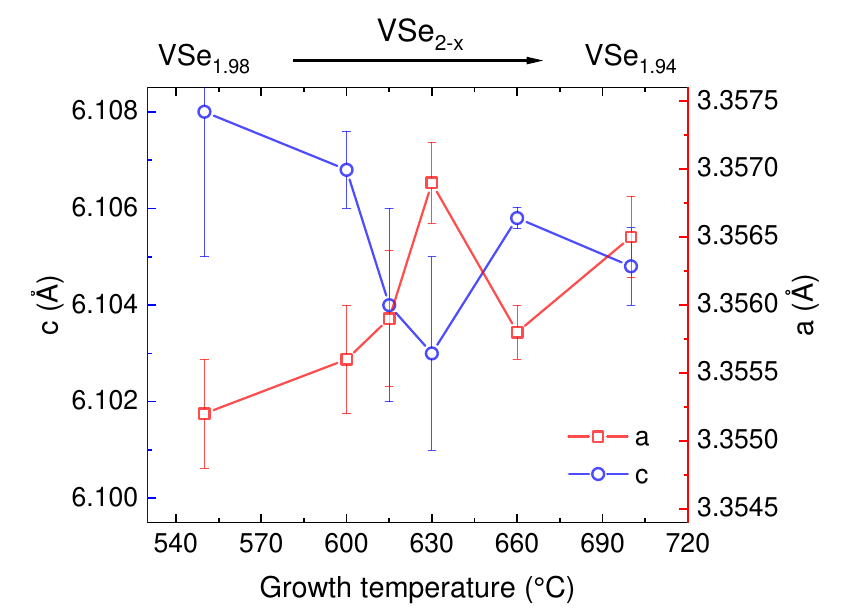}
	\end{center}
	\vspace*{-0.2in}
	\caption{\label{fig:PXRD} Lattice parameters of \VSe~samples obtained from powder x-ray diffraction (PXRD). The left and right plot axes show the \textit{c}- and \textit{a}-axis parameters of the P$\overbar{3}$m1 unit cell respectively. The top-axis labels shows a comparison with the overall variation of crystal stoichiometry (Se/V ratio) obtained from laboratory XPS measurements.}
\end{figure}

To provide further insight into the composition of our samples and to help elucidate the nature of defects, we obtained the crystal stoichiometry from independent lab-based XPS measurements (non-monochromatic Al-k$\alpha$ source, \hv~= 1486.7 eV) and lattice parameters from powder x-ray diffraction (PXRD) described in Ref.~\cite{Supp}. Shown in Fig.~\ref{fig:PXRD} are the results for the \textit{a}- and \textit{c}-axis lattice parameters as a function of growth temperature together with the variation of crystal stoichiometry for comparison (top-axis). Firstly, it can be seen that the stoichiometry (Se/V ratio) of our samples varies from Se/V = 1.98$\pm$0.10 to 1.94$\pm$0.10 with increasing growth temperature which is consistent with the electron-doping hypothesis (either V-rich or Se-deficient). Considering these two scenarios, it is important to first examine changes in the \textit{c}-axis parameter. Overall we find that it decreases with increasing growth temperature, and therefore it is unlikely that there are significant vanadium interstitials between the layers which may be expected to produce the opposite effect i.e. an \textit{increase} in the \textit{c}-axis by expanding the layer distance \cite{Morosan2006}. This is supported by our XPS and XAS data (Ref.~\cite{Supp}) that show no additional features which could be linked to extrinsic V species. Hence, we exclude the V-rich (V$_{1+x}$Se$_{2}$) scenario which would otherwise manifest as V interstitials in the vdW gap \cite{DiSalvo1976,DiSalvo1981}. Instead, we consider what would be the effects of Se-deficiency on the crystal structure. A study of controlled Se loss by annealing in the related compound, \TiSe~\cite{Huang2017} previously showed a decrease in the $c$-axis related to the Ti-Se bond length, as $\delta$ increases in TiSe$_{2-\delta}$ and occurs continuously across the range 350 - 950\degsC. At the same time, a corresponding increase in the a-axis was observed. This trend of the \textit{a}- and \textit{c}-lattice parameters is very similar to what we find for \VSe~in Fig.~\ref{fig:PXRD}. Therefore, we suggest that our results can be explained by an overall Se-deficiency (VSe$_{2-x}$) at higher growth temperatures due to increasing Se vacancies (defect B in Fig.~\ref{fig:Growth}). The desorption of Se at elevated temperatures is in line with previous reports \cite{Hildebrand2014,Peng2015} and seems to be of greater significance above 600-700\degsC~\cite{Zelenina2011}, which is consistent with our observations. As Se becomes volatile and leaves behind a vacancy, the majority is lost from the crystal. However, a small proportion may become trapped in-plane as aggregates/clusters or between layers as interstitials (defect C and E in Fig.~\ref{fig:Growth} respectively). This may explain the additional \SeMinus~selenium species seen in XPS [Fig.~\ref{fig:Se-3d}(a)]. It could also be possible that some additional Se is incorporated into the crystals due to the Se-rich environment during the growth process. This would likely occur at higher growth temperatures considering the greater energy cost to disrupt the lattice by introducing an aggregate or interstitial compared to a vacancy.

\begin{figure}[t!]
	\begin{center}
		\includegraphics[width=1.0\linewidth,clip]{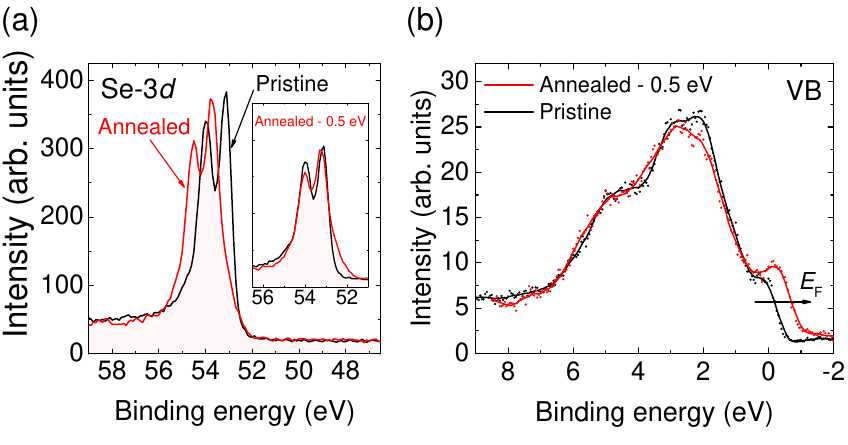}
	\end{center}
	\vspace*{-0.2in}
	\caption{\label{fig:annealingXPS} Effect of annealing on core levels measured by $\mu$XPS (\hv~= 1486.7 eV). (a) Se-3\textit{d} spectra for the \Tgrowth~= 630\degsC~sample. A comparison is made between the pristine sample (black) after cleaving in UHV, and the effect of annealing at 520\degsC~for $\sim$40 minutes (red). The inset shows the same comparison with a shift of -0.5 eV applied to the annealed sample spectrum. (b) Valence band (VB) spectra for the pristine and annealed sample (with -0.5 eV shift applied).}
\end{figure}

Finally, to mimic the effects of growing crystals at high temperatures, we also performed an annealing experiment to investigate the effect on the Se-3\textit{d} core levels and valence band (VB). Here, we used a $\mu$XPS technique with a focussed beam spot of 100 $\mu$m in order to precisely measure the same region of the sample before and after annealing. Shown in Fig.~\ref{fig:annealingXPS} are the results which compare the pristine sample (cleaved in ultra-high vacuum, UHV), with that following heat treatment (annealing to 520\degsC~for $\sim$40 mins). Similar to Ref.~\cite{Huang2017}, we expect annealing to result in Se loss from the sample, leading to a binding energy shift as the sample becomes overall electron doped. Indeed, this effect is clearly visible in Fig.~\ref{fig:annealingXPS}(a) and the inset shows that the core levels are shifted by approximately 0.5 eV. Although this is much larger than the intrinsic shift in Fig.~\ref{fig:Se-3d}(c), the annealing experiments were performed in UHV ($<$\sn{1}{-9} mbar) as opposed to the Se-rich atmosphere during crystal growth. Given the high vapor pressure of Se (0.07 atm at 520\degsC), the greater loss of Se in UHV is expected. As discussed previously, the overall electron doping raises the Fermi level meaning that the core levels shift to higher binding energy. This is confirmed by analyzing the VB spectra in Fig.~\ref{fig:annealingXPS}(b) where the same 0.5 eV shift is applied to the annealed sample spectrum such that the main features of the VB are overlaid. The up-shifted Fermi level position is indicated by the arrow. These results provide further evidence for the electron-doping scenario, specifically due to an overall Se-deficiency (VSe$_{2-x}$).

In summary, there are likely to be two major types of defects in \VSe~grown by CVT; namely Se vacancies and trapped Se aggregates or interstitials. Both will contribute to the defect scattering of carriers in electronic transport and hence decrease the RRR. However, the effective binding energy shift in XPS suggests that the dominant defect type is Se vacancies as the system becomes overall electron doped. This is further supported by the Se deficiency from measurements of the crystal stoichiometry, lattice parameters and annealing experiments. We suggest that the suppression of the CDW order is caused by a deviation from stoichiometry in the form of electron doping which reduces \Tcdw, similar to \TiSe~\cite{DiSalvo1976}. Although the density of defects and doping level are coincidentally linked, we suggest that the CDW is mainly influenced by doping. In fact, a recent STM investigation of the local density of states in the presence of defects showed the CDW gap in \VSe~to be extremely robust to disorder \cite{Jolie2019}. From our results, this is evidenced by the fact that the \Tcdw~approaches a maximum as the crystal stoichiometry becomes near 2:1 ratio at the lowest growth temperature concomitant with a saturation of the binding energy shift in XPS (doping level). Instead, the crystal purity (RRR) is expected to continue to increase below \Tgrowth~= 550\degsC~without any significant gain in \Tcdw. We note that this is approaching the lower limit of reported growth temperatures for 1\textit{T}-VSe$_2$ and other TMDs using the iodine CVT technique {\cite{Levy1979}}. By decreasing the temperature further, we anticipate that the growth rate will become increasingly slow until the required growth time is impractical or there is insufficient energy for crystallization to occur.

\section{Conclusion}\label{Sec:Conclusion}

From electronic transport measurements, we showed that both the CDW transition temperature, \Tcdw~and the residual resistance ratio (RRR) decreases for \VSe~crystals grown at higher temperatures. This is a result of an increased density of defects, which primarily manifests as Se vacancies as a result of Se desorption from the final crystal products at elevated temperature. Therefore, we suggest \VSe~becomes overall Se deficient which is confirmed by stoichiometry analysis and changes in the unit cell parameters from x-ray diffraction. This deficiency leads to an effective electron doping and a resulting up-shift in the Fermi level, which is evidenced by a rigid shift of Se-3\textit{d} core levels to higher binding energy in XPS measurements. Although the crystal purity is reduced at higher growth temperatures (lower RRR) due to a larger density of defects, we suggest that the effective electron doping due to overall Se-deficiency is mostly responsible for the reduction in \Tcdw. For samples with \Tgrowth~$<$~630\degsC, we see an increase in the CDW transition temperature above the typically reported value of \Tcdw~$\approx$110 K. Based on this knowledge, we were able to optimize the chemical vapour transport method in order to produce ultra-high purity \VSe~samples with near ideal stoichiometry at \Tgrowth~= 550\degsC~with \Tcdw~= (112.7 $\pm$ 0.8) K and maximum RRR $\approx$ 49. 

Our work highlights the importance of carefully controlling the growth conditions of strongly correlated TMDs whose properties are sensitive to defects and deviation from stoichiometry. In addition, the growth of high purity TMD materials is imperative for understanding both the bulk behaviour of these compounds and to provide a suitable source for exfoliation of monolayers and the preparation of van der Waals heterostructures.


\section*{Acknowledgements}\label{Sec:Acknowledgements}

We thank the Elettra synchrotron for access to the BEAR beamline (proposal no.~20180358) and the BEAR staff for assistance during measurements including S. Nannarone and N. Mahne for helpful discussions. Extensive technical support from P.~ Jones at the University of Bath is gratefully appreciated. We would like to thank S. Cross and S. Friedemann for insightful discussions and sharing preliminary data. We thank G. Balakrishnan for sharing unpublished information with regards to Ref.~\cite{Barua2017}. The authors acknowledge funding and support from the EPSRC Centre for Doctoral Training in Condensed Matter Physics (CDT-CMP), Grant No.\ EP/L015544/1. Finally, we acknowledge the Bristol NanoESCA Facility (EPSRC Strategic Equipment Grant No. EP/K035746/1 and EP/M000605/1).


\bibliographystyle{unsrtnat}


\pagebreak
\onecolumngrid
\begin{center}
	\textbf{\large Supplemental material for: ``Correlation between crystal purity and the charge density wave in 1\textit{T}-VSe$_{2}$"}
\end{center}

\setcounter{equation}{0}
\setcounter{figure}{0}
\setcounter{table}{0}
\setcounter{section}{0}
\setcounter{page}{1}
\makeatletter
\renewcommand{\thefigure}{S\arabic{figure}}
\renewcommand{\thetable}{S\arabic{table}}

\section{Summary of sample properties}\label{Sec:Summary of sample properties}

\begin{table*}[ht]
	\setlength{\tabcolsep}{7pt}
	\centering
	\begin{tabular}{c c c c c c c c}
		\hline\hline
		\Tgrowth~(\degsC) & \Treact~(\degsC) & \Tcdw~(K) & RRR range & Comp.
		& \multicolumn{2}{c}{Lattice const. (\AA)} \\ [0.5ex]
		& & & & & \textit{a} & \textit{c} \\ [0.5ex]
		\hline
		700 & 755 & 104.5 $\pm$ 0.9 &  10 - 13 & VSe$_{1.94 \pm 0.10}$ & 3.3556(3) & 6.1048(8)  \\
		660 & 720 & 108.1 $\pm$ 1.0 &  12 - 15 & -- & 3.3558(2) & 6.1058(2) \\
		630 & 690 & 101.7 $\pm$ 0.7 &  10 - 16 & VSe$_{2.25 \pm 0.10}$ & 3.3569(3) & 6.103(2) \\
		615 & 670 & 111.2 $\pm$ 0.6 &  19 - 28 & VSe$_{1.94 \pm 0.10}$ & 3.3559(5) & 6.104(2) \\
		600 & 655 & 112.3 $\pm$ 0.6 &  25 - 31 & VSe$_{1.97 \pm 0.10}$ & 3.3556(4) & 6.1068(8) \\
		550 & 610 & 112.7 $\pm$ 0.8 &  44 - 49 & VSe$_{1.98 \pm 0.10}$ & 3.3552(4) & 6.108(3) \\ [1ex]
		\hline
	\end{tabular}
	\caption{\label{tab:Summary of sample properties} Summary of sample properties including; growth temperature (T$_{g}$), reaction temperature (T$_{r}$), CDW transition temperature (\Tcdw), range of RRR observed within the batch, chemical composition (Comp.) from XPS measurements, and unit-cell lattice constants from powder x-ray diffraction (PXRD).} 
\end{table*}

\noindent
Presented in Table.~\ref{tab:Summary of sample properties} is a summary of the electronic and physical properties of the sample batches grown at different temperatures. Determination of the electronic properties including RRR and transition temperature, \Tcdw~is discussed in the main text and in supplementary section \ref{Sec:Determining the CDW transition temperature}.

In order to determine the overall composition of our samples, additional XPS measurements were performed at the Bristol NanoESCA Facility. An unpolarized, non-monochromatic Al-k$\alpha$ (\hv~= 1486.7 eV) source was used together with an ARGUS analyser at 45$^{\circ}$ with respect to the normal of the sample, and the pass energy was 50 eV. The stoichiometry was calculated using the entire area of Se-3\textit{d} and V-2\textit{p} photoemission lines, applying a Shirley background subtraction and normalizing the spectra by considering the specific sensitivity factor and inelastic mean-free path. The Se/V ratio of different samples and the composition of VSe$_{x}$ is shown in Table.~\ref{tab:Summary of sample properties}. We believe that the stoichiometry of the 630\degsC~sample is an anomalous result if we compare to trends of the electronic transport and synchrotron XPS in the main text. The crystal available for these measurements was particularly small and it is possible that, despite our best efforts, the surface was only partially cleaved and/or the results are skewed by the overall reduced photoelectron intensity from the smaller sample.

To determine the crystal lattice constants, powder x-ray diffraction (PXRD) measurements were carried out on a STOE STADI P instrument using Ge-monochromated Cu-K$\alpha$1 radiation ($\lambda$ = 1.54060 \AA) operating in transmission mode. The scan range was (20-93)$^{\circ}$ in 2$\theta$. Two Dectris 1K Mythen detectors were used to record data.

The \VSe~crystals were plate-like and even after grinding into a powder, they showed preferred orientation. For peak indexing and refinement of the unit cell, the Werner \cite{suppWerner1985} or Lou\"{e}r \cite{suppBoultif2004} algorithms implemented in WINXPOW \cite{suppWINXPOW2018} were used. All samples were successfully indexed to the trigonal (P$\overbar{3}$m1) unit cell expected for \VSe~\cite{suppEaglesham1986}. The resulting lattice parameters are given in Table. \ref{tab:Summary of sample properties}

\section{Determining the CDW transition temperature}\label{Sec:Determining the CDW transition temperature}

\begin{figure}[ht]
	\floatbox[{\capbeside\thisfloatsetup{capbesideposition={right,center},capbesidewidth=6cm}}]{figure}[\FBwidth]
	{\includegraphics[width=8.35cm]{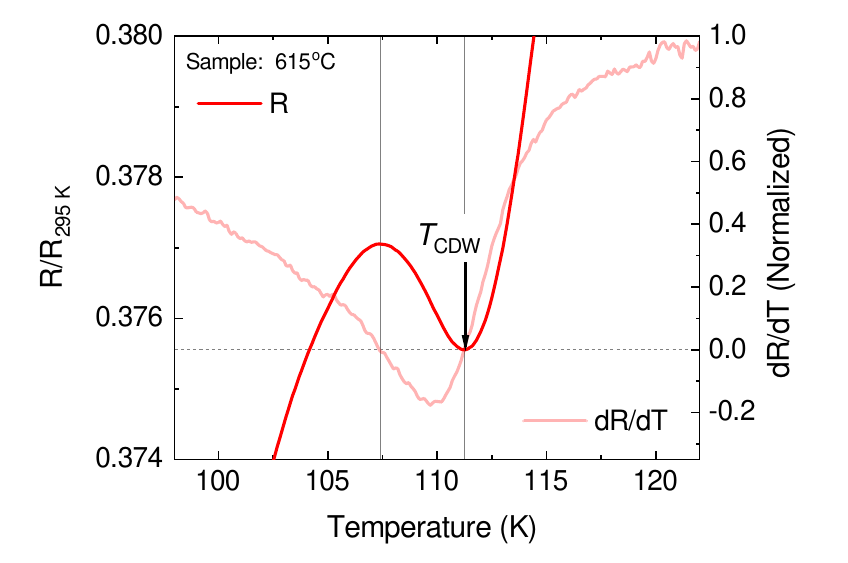}}
	{\caption{\label{fig:Transitions1} Determining the CDW transition temperature from resistance measurements. \Tcdw~is the minimum before the inflexion point where the resistance (left axis) starts to increase near $\sim$110 K, which also corresponds to point at which \dRdT~(right axis) crosses zero.}}
\end{figure}

\noindent
As discussed in the main text, the first derivative of resistance, \dRdT~highlights the CDW transition in \VSe. We define \Tcdw~as the minimum before the inflexion point where the resistance, \textit{R} begins to increase near $\sim$110 K as indicated by the arrow in Fig.~\ref{fig:Transitions1}(a). This corresponds to the point at which \dRdT~initially crosses zero, which is \Tcdw~= (111.2 $\pm$ 0.6) K for the 615\degsC~example. The error has been obtained by evaluating the uncertainty in the \dRdT~= 0 crossing based on a linear fit to the steep decline before \Tcdw. The same methods were used to define \Tcdw~for all samples in the main text.

Here, we also highlight the two vertical solid lines corresponding to the zero crossing points of \dRdT. These points also coincide with the minimum and maximum of the increase in resistance as a result of the CDW transition. In this region, \dRdT~becomes negative, and its magnitude below zero therefore provides an indication of the insulating state (resistance increase) that develops near \Tcdw~which is related to the proportion of carriers removed from the Fermi surface at the CDW transition.

\section{Temperature dependence of the resistance}\label{Sec:Temperature dependence of the resistance}

\begin{figure}[h]
	\begin{center}
		\includegraphics[width=0.88\linewidth,clip]{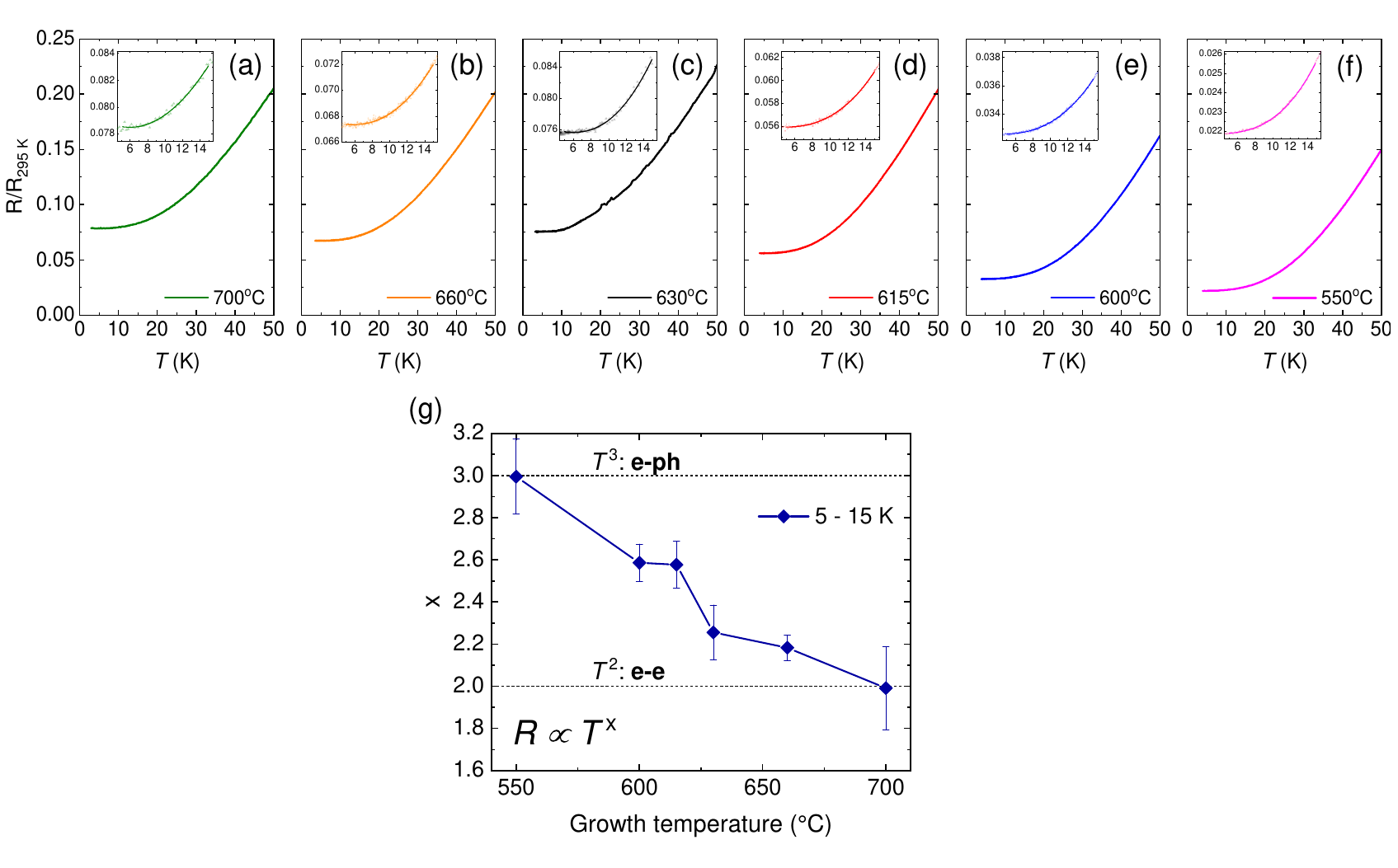}
	\end{center}
	\vspace*{-0.2in}
	\caption{\label{fig:Power} Temperature dependence of resistance. (a) - (e) Normalized resistance for all samples in the range (4 - 50) K. The inset shows a power law fit, R $\propto T^{x}$ to the low temperature data (5 - 15 K). (f) x in  R $\propto T^{x}$ as a function of growth temperature. The errors bars are related to the fitting only. The dashed lines indicate the $T^{2}$ (electron-electron) and $T^{3}$ (electron-phonon) carrier scattering regimes respectively.}
\end{figure}

\noindent
By fitting the low temperature resistance data (5 - 15 K) with a power law it is possible to extract information on the dominant scattering mechanism of electrons. Shown in Fig.~\ref{fig:Power}(a) - (e) is the normalized resistance $<$ 50 K for all samples plotted using the same scale. The increasing residual resistance discussed in the main text is obvious here at higher growth temperatures. The insets shows the data ($<$ 15 K) with a power law fit. Fig.~\ref{fig:Power}(f) shows the result of the fitting for \textit{x} in R $\propto T^{x}$ as a function of growth temperature. The sample grown at 700\degsC~(lowest purity, avg. RRR $\approx$ 11.5) shows a $T^{2}$ dependence indicating an electron-electron dominated carrier scattering, which has been noted before in this compound \cite{suppThompson1979}. For samples grown at lower temperatures, \textit{x} progressively increases until it reaches a $T^{3}$ dependence for the 550 \degsC~sample (highest purity, avg. RRR $\approx$ 46.5) with near ideal stoichiometry (VSe$_{1.98}$). A $T^{3}$ dependence has been linked to electron-phonon dominated scattering in other TMDs \cite{suppNaito1982}. Interestingly, this evolution seems similar to the trend found for \Tcdw~and RRR in the main text which was linked to the doping level introduced by defects.

\section{X-ray spectroscopy of vanadium core levels}\label{Sec:XPS results of V-2p}

\begin{figure*}[ht]
	\begin{center}
		\includegraphics[width=0.9\linewidth,clip]{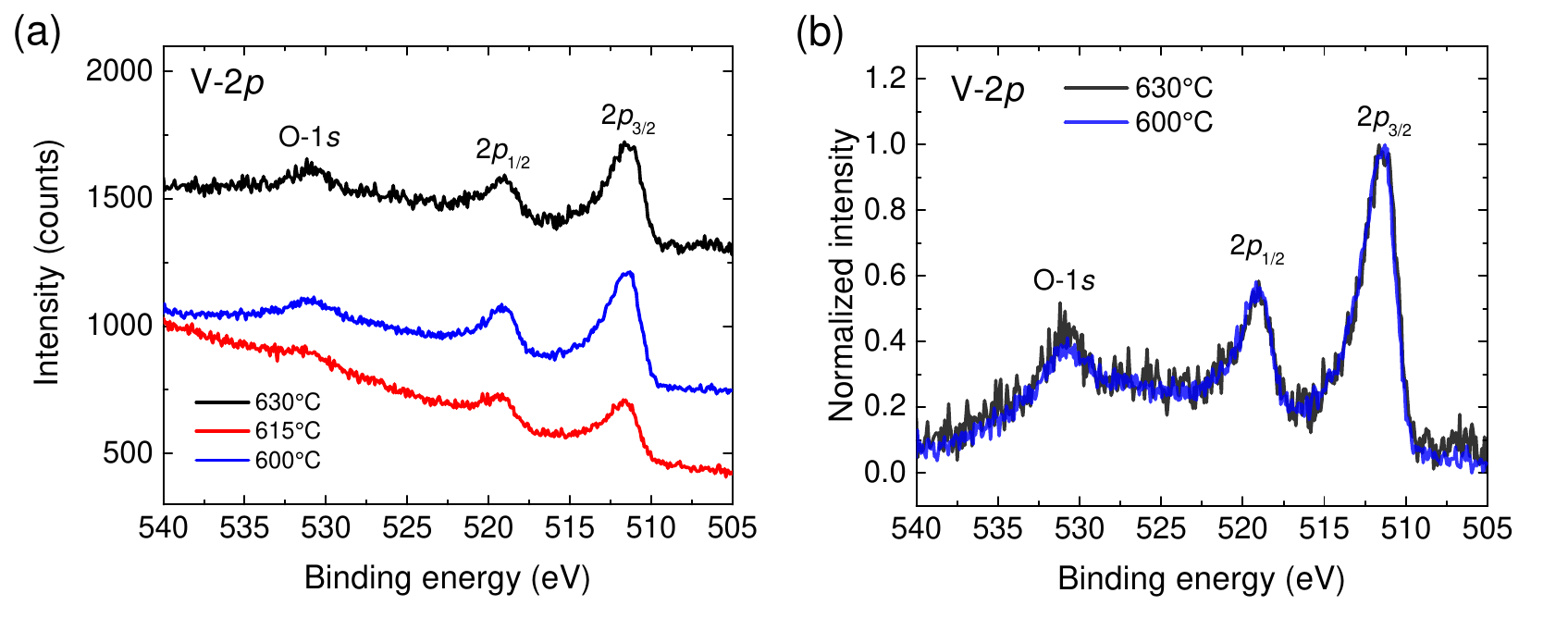}
	\end{center}
	\vspace*{-0.2in}
	\caption{\label{fig:XPS_V-2p} XPS spectra of V-2\textit{p} core levels (\hv~= 630 eV). (a) Raw data showing the doublet consisting of V-2\textit{p}$_{1/2}$ and V-2\textit{p}$_{3/2}$ (b) A direct comparison of the samples with contrasting electronic properties (\Tgrowth~= 630 and 600\degsC~with avg. RRR $\sim$13 and $\sim$28 respectively). The data has been normalized to the V-2\textit{p}$_{3/2}$ peak, after subtraction of a Shirley background.}
\end{figure*}

\noindent
In addition to the Se-3\textit{d} spectra presented in the main text, we obtained additional measurements of the V-2\textit{p} core levels for 3 samples with a range of growth temperatures over which the electronic properties are changing significantly \Tgrowth~= (600 - 630)\degsC. Shown in Fig.~\ref{fig:XPS_V-2p}(a) is the raw V-2\textit{p} XPS spectra. We find the doublet consisting of V-2\textit{p}$_{1/2}$ and V-2\textit{p}$_{3/2}$ at 519.1 eV and 511.3 eV respectively ($\Delta_{so}$ = 7.8 eV). Additionally, there is a small oxygen (\textit{O-1\textit{s}}) feature at $\sim$530.7 eV which we believe is not intrinsic to the crystals (see later discussion of XAS data). Our results are qualitatively similar to previous studies of V-2\textit{p} levels in \VSe~\cite{suppBonilla2018,suppLiu2018,suppCao2017,suppGhobadi2017}, although there seems to be large variation in the literature. Overall, we find our results are in closest agreement with Ref.~\cite{suppBonilla2018} if we consider the position of O-1\textit{s} in that data (531.3 eV) as a reference.

There is no significant variation of the V-2\textit{p} spectra between our samples as we find the binding energies and lineshape of all features to be nearly identical. This is particularly evident in Fig.~\ref{fig:XPS_V-2p}(b) in which we show the normalized data for the samples with the most contrasting CDW properties (\Tgrowth~= 630 and 600$^{\circ}$C). In addition, we did not observe any features would could be linked to extrinsic vanadium species which has previously been ascribed to interstitials in the van der Waals gap occurring at higher growth temperatures \cite{suppDiSalvo1981}. Therefore, we conclude that there are no significant variations in the vanadium (V-2\textit{p}) states that can explain the observed difference in CDW properties between our samples.

\begin{figure*}[ht]
	\begin{center}
		\includegraphics[width=0.5\linewidth,clip]{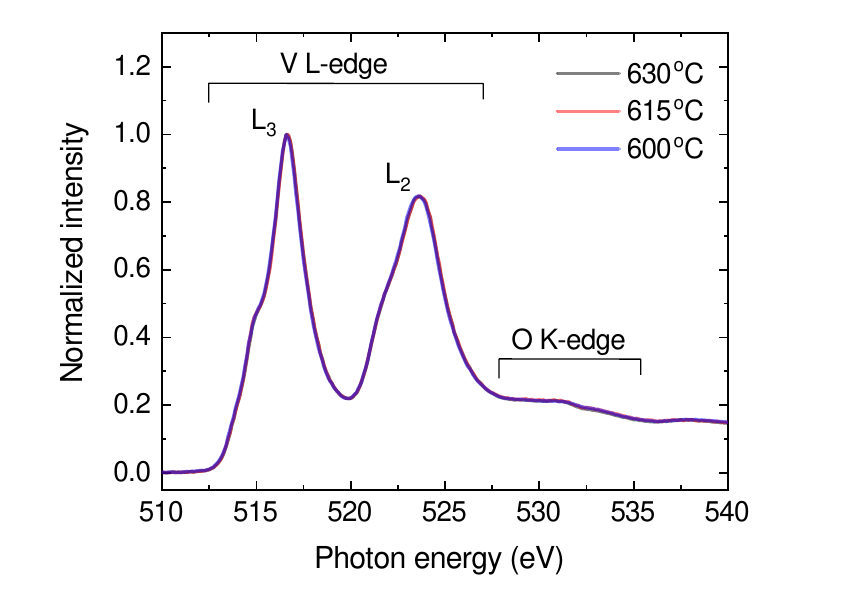}
	\end{center}
	\vspace*{-0.2in}
	\caption{\label{fig:XAFS} X-ray absorption spectroscopy (XAS) near the V L-edge and O K-edge. The spectra have been normalized to the maximum of the V L$_{3}$-edge.}
\end{figure*}

To further examine the vanadium states, we also performed X-ray absorption spectroscopy (XAS) measurements of the V L-edge using the total electron yield (TEY) method of monitoring the drain current on the sample as a function of the incident photon energy. The results are shown in Fig.~\ref{fig:XAFS}. The main two features at 516.6 eV and 523.6 eV are related to excitations of the V-2\textit{p}$_{3/2}$ (L$_{3}$-edge) and V-2\textit{p}$_{1/2}$ (L$_{2}$-edge) core levels into unoccupied V-3d states respectively \cite{suppBiener2000}.

The XAS spectra for all samples are nearly indistinguishable, both near the V L-edge and O K-edge regions. By examining the O K-edge in these spectra, we suggest that the amount of oxygen throughout our samples is negligible. Instead, the O-1\textit{s} detected by XPS (Fig. \ref{fig:XPS_V-2p}) was likely only present on the surface or in the local environment such as the measurement chamber or sample holder, and not intrinsic to the crystals. Therefore, we rule out the presence of oxygen as a possible explanation for the observed difference in CDW properties of our samples.

\section{x-ray spectroscopy of iodine core levels}\label{Sec:XPS results of I-3d}

\noindent
Since iodine is used as the transport agent in the CVT growth process, we looked in the spectral region where we woudl expect I-3\textit{d} core levels to see if iodine had inadvertently been incorporated into our samples. This effect has been documented in the related compound, \TiSe~in which approx. 0.3 at.\% iodine was found regardless of the growth temperature \cite{suppDiSalvo1976} in the range \Tgrowth~= (600 - 900)\degsC. In that study, it was found that samples grown using iodine had a slightly reduced CDW transition compared to those grown by direct sublimation in excess Se i.e. without iodine.

\begin{figure*}[ht]
	\begin{center}
		\includegraphics[width=0.5\linewidth,clip]{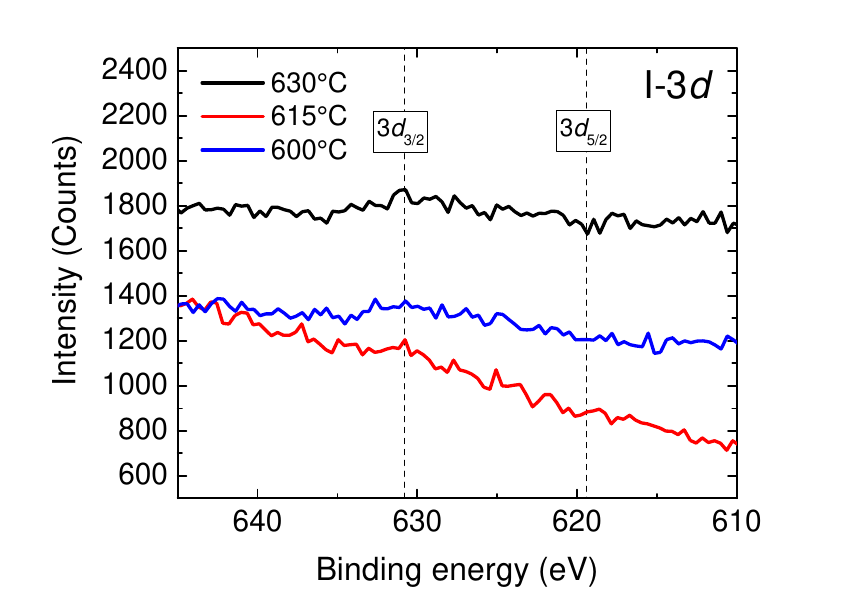}
	\end{center}
	\vspace*{-0.2in}
	\caption{\label{fig:XPS_I-3d} XPS spectra of I-3\textit{d} core levels (\hv~= 730 eV). The raw data is shown. The vertical dashed lines indicate the expected peak positions of the I-3\textit{d}$_{5/2}$ and I-3\textit{d}$_{3/2}$ doublet.}
\end{figure*}

Presented in Fig.~\ref{fig:XPS_I-3d} is the raw XPS data in the spectral region where we expect I-3\textit{d} core levels. A doublet is expected consisting of I-3\textit{d}$_{5/2}$ and I-3\textit{d}$_{3/2}$ at 619.3 and 630.8 eV respectively [vertical dashed lines in Fig.~\ref{fig:XPS_I-3d}]. Based on our data, it is possible to speculate that there is a weak feature in the region (620 - 635) eV but its magnitude is comparable to the noise. At \hv~= 730 eV photon energy, the photoionization cross-section of I-3\textit{d} should be $\sim$ 2.7 Mbarn, which is more than double the cross-section of V-2\textit{p} (at \hv~= 630 eV) which was clearly detectable above the noise in our measurements (see Fig. \ref{fig:XPS_V-2p}). Therefore, we suggest that iodine is either not present in our samples or it is below the detectable limit.

Moreover, it can be seen that the samples with most contrasting electronic properties (\Tgrowth~= 630 and 600$^{\circ}$C) have nearly identical spectra. Therefore, we can say that if there is iodine present in the crystals, either the concentration does not change significantly with growth temperature and/or it has little effect on the observed CDW properties.

\end{document}